\documentclass[12pt]{article}
\usepackage{epsfig}
\usepackage[hang,bf,small]{caption}
\DeclareGraphicsExtensions{.eps.gz,.eps,.ps,.ps.gz}
\parskip=\itemsep               %?
\begin{document}
\begin{titlepage}
\noindent
\begin{flushright}
\parbox[t]{10em}{
DESY 99-113 \\
TAUP  2594 - 99\\
 \today\\
%{\tt hep-ph/9908???}
}
\end{flushright}
\vspace{1cm}
\begin{center}
{\Huge  \bf   Screening     Effects  on  $\mathbf{F_2}$ }\\
{\Huge \bf at Low $\mathbf{x}$ and $\mathbf{Q^2}$}
 \\[4ex]
 
{\large \bf  E. ~L e v i n $\mathbf{{}^{1)}}$}\\[4.5ex]

\footnotetext{$^{1)}$ Email: leving@post.tau.ac.il .}

{\it  HEP Department}\\
{\it  School of Physics and Astronomy}\\
{\it Raymond and Beverly Sackler Faculty of Exact Science}\\
{\it Tel Aviv University, Tel Aviv, 69978, ISRAEL}\\
{\it and}\\
{\it DESY Theory Group}\\
{\it 22603, Hamburg, GERMANY}\\[4.5ex]
\end{center}
~\,\,
\vspace{1cm}

{\em Talk, given at Ringberg Workshop: ``New Trands in HERA Physics", May
30 - June 4 ,1999}

\vspace{1cm} 
\samepage
{\large \bf Abstract:}
In this talk we discuss  how deeply  the region of high parton densities
has been studied experimentally at HERA. We show that the measurements of
deep inelastic structure functions at HERA confirm our theoretical
expectation that at HERA we face a challenging problem of understanding a
new system of partons: quarks and gluons at short distances  with so
large densities that we cannot treat this system perturbatively.
We collect all experimental indications and manifestations of specific
properties of high parton density QCD.

\end{titlepage}

\section{What Are Shadowing Corrections?}
In the region of low $x$ and low $Q^2$ we face two challenging problems
which have to be resolved in QCD:
\begin{enumerate}
\item\,\,\, Matching of ``hard" processes, that can be successfully 
described in perturbative QCD (pQCD),  and ``soft" processes, that should
be described in non-perturbative QCD (npQCD), but actually, we have only a
phenomenological approach for them;

\item\,\,\, Theoretical approach for the high parton  density QCD (hdQCD) 
which we reach in the deep inelastic scattering at low $x$ but at
sufficiently high $Q^2$.  In this kinematic region we expect that the
typical distances will be small but the parton density will be so large
that a new non perturbative approach shall be developed for understanding
this system.
\end{enumerate} 
 
We are going to advocate the idea that these two problems are correlated
and the system of partons always passes the stage of hdQCD before
( at shorter distances ) it goes to the black box, which we call
non-perturbative QCD and which, practically, we describe in old fashion 
Reggeon phenomenology. In spite of the fact that there are many reasons to 
believe that such a phenomenology could be even correct, the distance
between Reggeon approach and QCD is so large we are loosing any taste of
theory doing this phenomenology. In hdQCD we still have a small parameter (
running QCD coupling $\alpha_S$ ) and we can start to approach this
problem using the developed methods of pQCD  \cite{GLR}. However, we
should
realize
that the kernel of the hd QCD problems are non-perturbative one, and 
therefore, approaching hdQCD theoretically we are preparing a new training
grounds for  searching methods for npQCD.   

First, let me recall that DIS experiment is nothing more than a microscope
and we have two variables to describe  its work. The first one is the
resolution  of the microscope, namely, $\Delta x \,\approx\,\,1/Q$ where
$Q^2$ is the virtuality of the photon.  It means that out microscope can
see  all constituents inside a target  with the size larger that $\Delta
x$. 
The second variable is time of observation. It sounds strange that we have 
this new variable, which we do not use,  working with a medical
microscope.
However, we are  dealing here with the relativistic system which can
produce hadrons (partons). So, for everyday analogy, we should consider
rather a box with flies which  multiply  and their number is, certainly,
different in different moment of time. To estimate this time we can use
the uncertainty principle $\Delta t \,\propto 1/\Delta E $ where $\Delta
E $ is the change of  energy, namely, $\Delta E = E_{initial}  -
E_{final}$, and for system of quark and antiquark $\Delta E = q_0 -
 q_1 - q_2 = q_0 - q = \frac{(q_0 - q) (q_0 + q)m}{2 q_0 m}
 = \frac{Q^2 m}{W^2} = mx$, where $m$ is mass of the target and $q_0$ and
$q$ is the energy and momentum of the virtual photon. Finally, $t = 1/mx$
with $x = \frac{Q^2}{W^2}$ where $W$ is the energy of photon - target
interaction.

Therefore, the question, that we are asking in DIS at low $x$, is what
happens with
constituents of rather small size after long time. It is clear that the
number of these constituents should increase since in QCD each parton can
decay in two partons with the   probability $P_i = \frac{N_c\alpha_S}{\pi}
\frac{d E_i}{E_i} \,\frac{d^2 k_{i,t}}{k^2_{i,t}}$ where $E_i$ and
$\vec{k}_i$  are energy and momentum of an emitted parton $i$.

This growth  we can describe introducing so called structure function 
( $xG(x,Q^2)$ )  or the number of partons that can be resolved with the
microscope with definite $Q^2$ and $x$.  Indeed,
\begin{equation} \label{DGLAP}
\frac{ \partial^2 x G(x,Q^2)}{\partial \ln(1/x) \,\partial
\ln Q^2}\,\,=\,\,\frac{N_c \alpha_S}{\pi} x G(x,Q^2)\,\,.
\end{equation}
This equation is the DGLAP\cite{DGLAP}      evolution equation in the
region
of low $x$. It has an obvious solution
 $xG(x,Q^2)
\,\propto
\,exp\left( 2 \sqrt{\frac{N_c
\alpha_S}{\pi}\,\ln(1/x)\,\ln(Q^2/Q^2_0)}\right)$.  
Therefore, we expect the increase of the parton densities at $x
\,\rightarrow 0$.

In Fig. 1 we picture the parton distributions in the transverse plane. At
$x \approx 1$ there are several partons of a small size. The distance
between partons is much larger than their size and we can neglect 
interactions between them.

\begin{figure}[hbtp]
\begin{center}  
\includegraphics[width=0.9\textwidth]{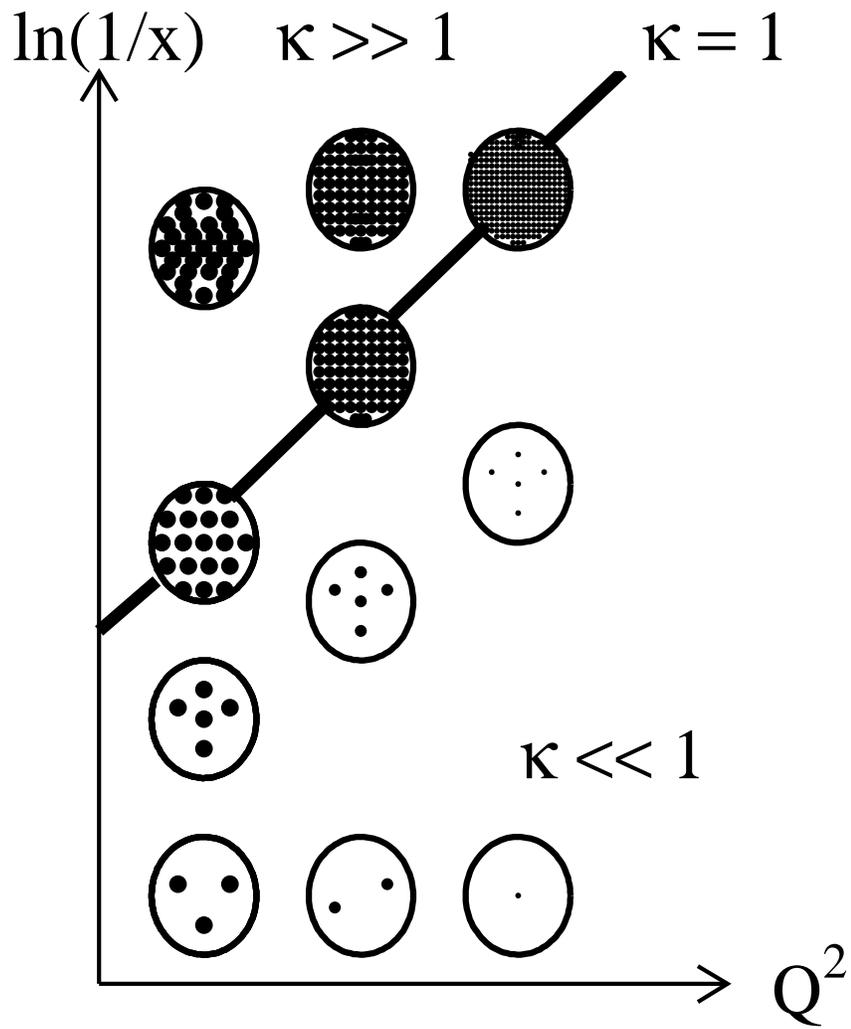}
\end{center}
\caption[]{Parton distribution in the transverse plane}
\label{eps1.1}
\end{figure}
 
However, at $x \,\rightarrow\,0$ the number of partons becomes so large
that they are populated densely in the area of a target. In this case, you
cannot neglect the interactions between them which was omitted in the
evolution equations ( see Eq. (~\ref{DGLAP} ) ). Therefore, at low $x$ we
have a more complex problem of taking into account both emission and
rescatterings of partons. Since the most important in QCD is the three
parton interaction, the processes of rescattering is actually a process of
annihilation in which one parton is created out of  two partons (gluons).  

Therefore, at low $x$ we have two processes
\begin{enumerate}
\item\,\,\, Emission induced by the QCD vertex $ G + G \rightarrow G$ with
 probability which is proportional to
$ \alpha_S \,\rho$ where $\rho$ is the parton density in the transverse
plane , namely
\begin{equation}
\rho  = \frac{ x G(x,Q^2 )}{\pi R^2}\,\,,
\end{equation}
where $\pi R^2$ is the target area;
\item\,\,\,Annihilation induced by the same vertex $ G + G \rightarrow G$
with  probability which is proportional to $\alpha_S \sigma_0
\,\rho^2$, where $ \alpha_S$ is probability of the processes $ G + G
\rightarrow G$,  $\sigma_0 $ is the cross section of two parton
interaction and $\sigma_0 \propto \,\frac{\alpha_S}{Q^2}$.  $\sigma_0
\,\rho$ gives the probability for two partons to meet and to interact,
while  $\alpha_S \sigma_0
\,\rho^2$ gives the probability of the annihilation process.
\end{enumerate}
 
Finally, the change of  parton density is equal to \cite{GLR} \cite{MUQI}
\begin{equation} \label{GLR}
\frac{ \partial^2 \rho(x,Q^2)}{\partial \ln(1/x) \,\partial
\ln Q^2}\,\,=\,\,\frac{N_c \alpha_S}{\pi} \rho(x,Q^2)\,\,-\,\,
\frac{\alpha^2_S\gamma}{Q^2}\, \rho^2(x,Q^2)\,\,
\end{equation}
or in terms of the gluon structure function
\begin{equation} \label{GLRG}
\frac{ \partial^2 x G(x,Q^2)}{\partial \ln(1/x) \,\partial
\ln Q^2}\,\,=\,\,\frac{N_c \alpha_S}{\pi} x G(x,Q^2)\,\,-\,\,
\frac{\alpha^2_S\gamma}{\pi R^2 Q^2}\,\left( x G(x,Q^2) \right)^2\,\,,
\end{equation}
where $\gamma$ has been calculated in pQCD \cite{MUQI}.

Therefore, Eq.(~\ref{GLRG}~) is a natural generalization of the DGLAP
evolution equations. The question arises,  why we call shadowing and /or
screening corrections  such a natural 
equation for a balance of partons due to two competing processes.
To understand this let us consider the interaction of the
fast hadron with the virtual photon at the rest ( Bjorken frame ).
In the parton model, only the slowest ( ``wee" ) partons interact with the 
photon. If the number of the ``wee" partons $N$  is not large,  the cross
section
is equal to $\sigma_0 N $. However, if we have two ``wee" partons with the
same energies and momenta, we overestimate the value of the total cross
section using the above formula. Indeed, the total cross section counts
only the number of interactions and, therefore, in the situation when one
parton is situated just behind another we do not need to count the
interaction of the second parton if we have taken into account the
interaction of the first one. It means that the cross section is equal to
\begin{equation} \label{SCGEN}
\sigma_{tot}\,\,=\,\,\sigma_0 N \,\{\,\,1\,\,-\,\,\frac{\sigma_0}{\pi R^2} 
\,\,\}\,\,,
\end{equation}
where $R$ is the hadron  radius. One can see that we reproduce
Eq.(~\ref{GLRG} ) by taking into account that there is a probability
for a parton not to interact being in a shadow of the second parton.

\section{What Have We Learned about SC?}
During the past two decades high parton density QCD has been under the
close investigation of many theorists \cite{GLR}
\cite{MUQI}\cite{THEOR}\cite{ML} and we summarize here
the result of their activity.

\begin{itemize}

\item\,\,\, The parameter which controls the
strength of SC has been found and it is equal to
\begin{equation} \label{KAPPA}
\kappa\,\,\,=\,\,\,\frac{3\,\pi^2 \alpha_s}{2
Q^2}\,\times\,
\frac{xG(x,Q^2)}{\pi\,R^2}\,\,\,=\,\,\,\sigma_0\,\times\,\rho(x,Q^2)\,\,.
\end{equation}
The meaning of this parameter is very simple. It gives the probability
of interaction for two partons in the parton cascade or, better to say, a
packing factor for partons in the parton cascade.
 
\item\,\,\,We know the correct degrees of freedom at high energies: colour
dipoles \cite{MUDIPOLE}. By definition, the correct degrees of freedom is
a set of quantum numbers which mark the wave function that is diagonal
with respect to interaction matrix. Therefore, we know that the size and
the
energy of the colour dipole are not changed by the high energy QCD
interaction.  

\item\,\,\, A new scale $Q^2_0(x)$ for hdQCD has been traced in pQCD
approach which is a solution to the equation
\begin{equation}\label{NEWSCALE}
\kappa\,\,\,=\,\,\,\frac{3\,\pi^2 \alpha_s}{2
Q^2_0(x)}\,\times\,
\frac{xG(x,Q^2_0(x))}{\pi\,R^2}\,\,=\,\,1\,\,.
\end{equation}
This new scale leads to the effective Lagrangian approach which 
gives us a general non-perturbative method to study hdCD.

\item\,\,\,We know that the GLR  equation  ( see Eqs.(~\ref{GLR}
) and (~\ref{GLRG}) ) describes the evolution
of the dipole density in the full kinematic region \cite{KOVCH}.
We understood that the Mueller-Glauber approach for colour dipole
rescattering  gives the initial condition to the GLR equation.

\item\,\,\, The new, non-perturbative approach, based on the effective
Lagrangian \cite{ML}, have been developed for hdQCD which gives rise to
the hope that hdQCD can be treated theoretically from the first
principles. 

\item\,\,\, We are very close to understanding  of the parton density
saturation \cite{GLR}.
\end{itemize}

In general, we think that the theoretical approach to hdQCD in a good
shape now.

\section{HERA Puzzle: Where Are SC?}
The wide spread opinion is that HERA experimental data for
$Q^2\,\geq\,1\,GeV^2$ can be described quite well using only the DGLAP
evolution equations, without any other ingredients such as shadowing
corrections, higher twist contributions and so on ( see. for example,
reviews \cite{EXPREV} ). On the other hand, the most important HERA
discovery is the fact that the density of gluons ( gluon structure
function ) becomes large in HERA kinematic region
\cite{EXPREV}\cite{ZEUSGL}.  The gluon densities extracted from HERA data
is so large that parameter $\kappa$ ( see Eq.(~\ref{KAPPA}) )  exceeds
unity
in substantial part of HERA kinematic region (see Fig.2a ). Another way to
see this is to plot the solution to Eq.(~\ref{NEWSCALE}) ( see Fig.2b).
It means that in large kinematic region $\kappa\,\geq\,1$ ( to the left
from line $\kappa = 1 $ in Fig.2b), we expect that the SC should be large
and important for a description of the experimental data. At first sight
such expectations are in a clear contradiction with the experimental data.
Certainly, this fact gave rise to the suspicions or even mistrusts that
our theoretical approach to SC is not consistent. However, the revision
and re-analysis of the SC , as has been discussed in the previous section,
have been completed with the result, that $\kappa$ is responsible for the
value of SC.

\begin{figure}[hbtp]
\begin{tabular} {c c }
\includegraphics[width=.5\textwidth]{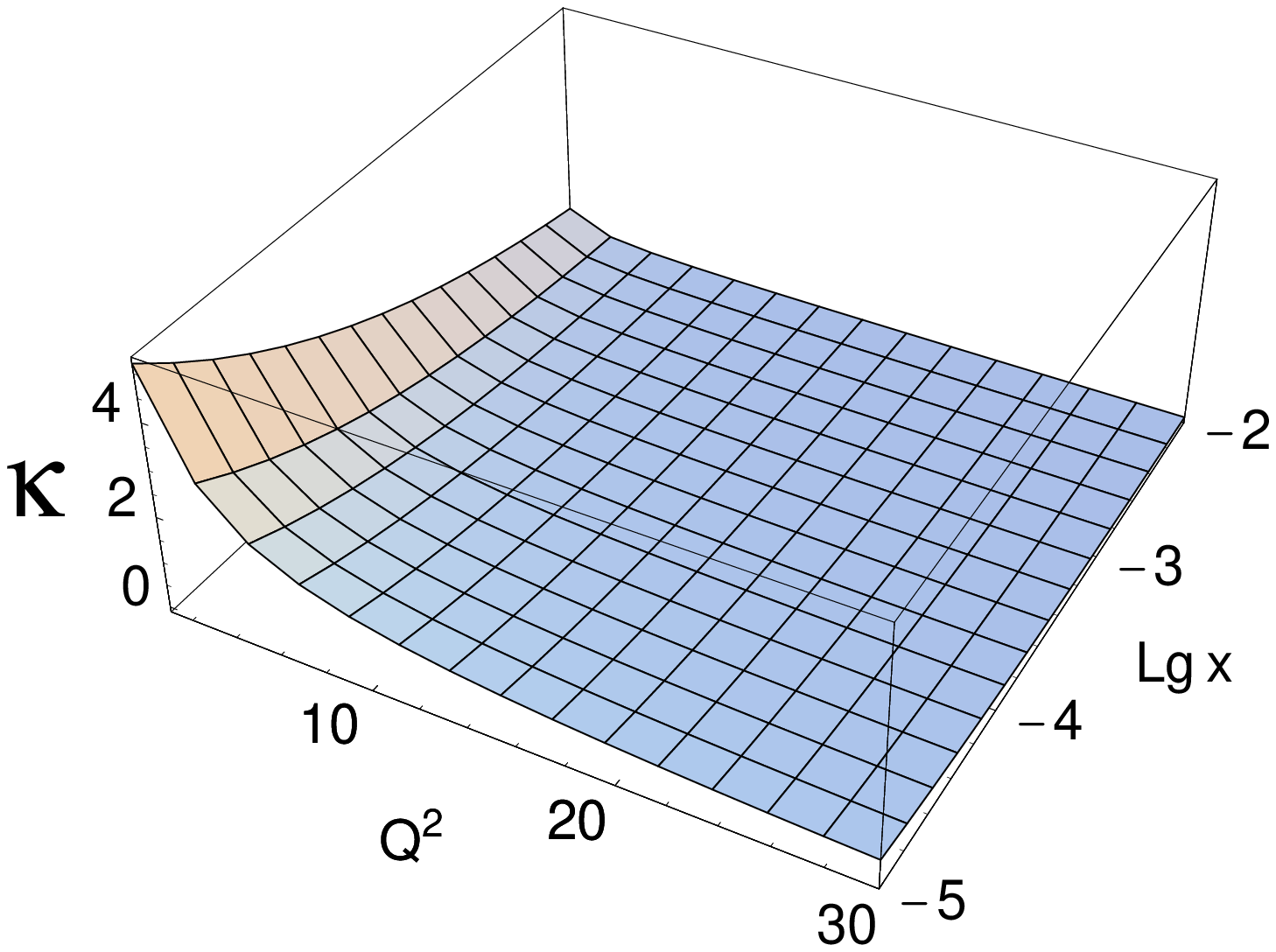} &
\includegraphics[width= 0.4\textwidth]{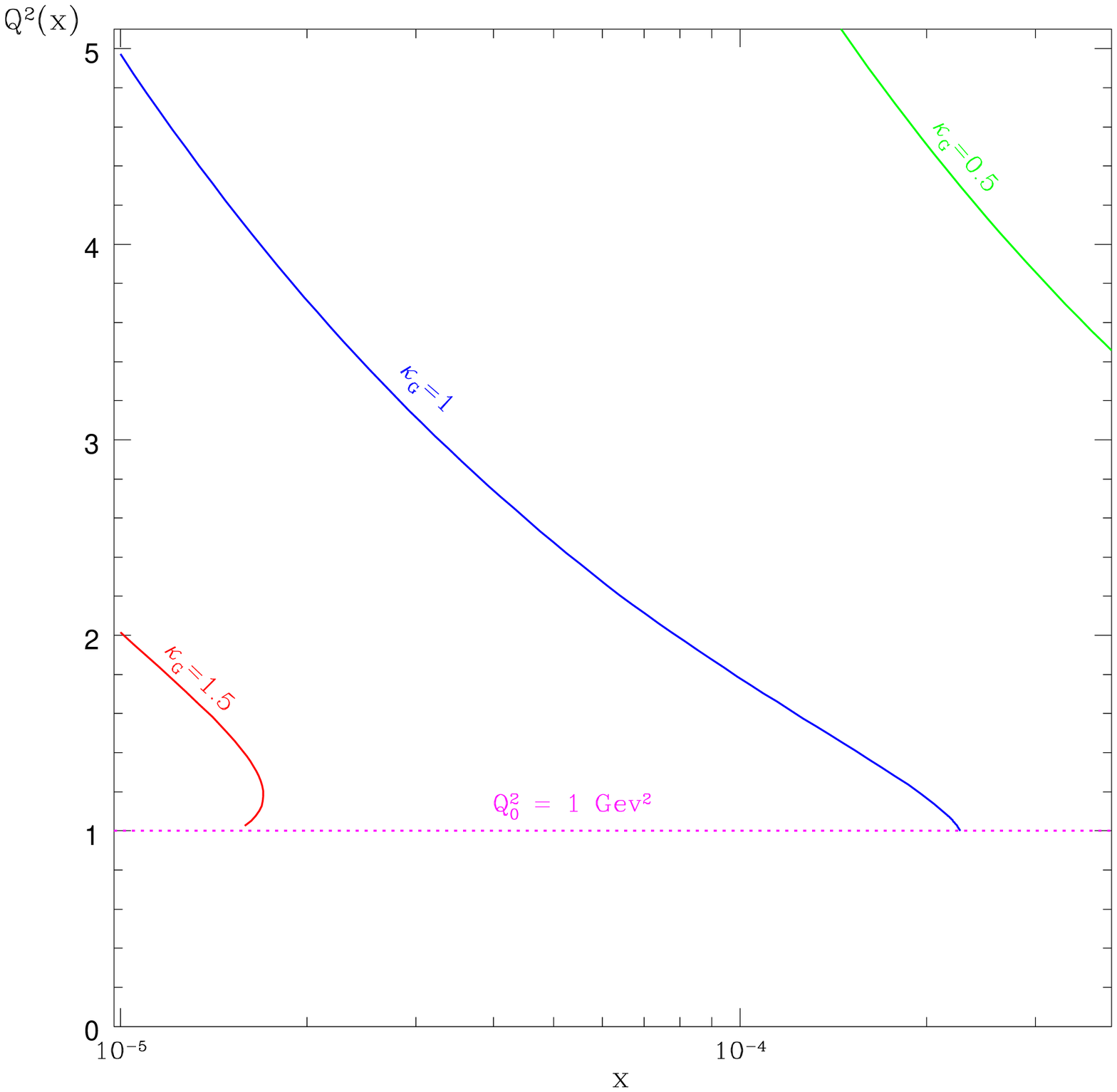}\\
Fig.2a & Fig.2b
\end{tabular}
\caption{Parameter $\kappa$ and new scale $Q^2_0(x)$.} 
\end{figure}

Therefore, we face a puzzling question: {\it where are SC?}. Actually,
this
question includes, at least, two questions: (i) why SC are not needed to
describe the HERA data on $F_2$, and (ii) where are the experimental
manifestation of strong SC. The answers for these two questions you will
find in the next three sections but, in short, they are: SC give a weak
change for $F_2$ in HERA kinematic region, but they are strong for the
gluon structure function . We hope to convince you that there are at
least two indications in HERA data supporting a large value of SC to  
gluon density:
\begin{enumerate}
\item\,\,\, $x_P$ - behaviour of the cross section of the diffractive
dissociation ($\sigma^{DD}$) in DIS;
\item\,\,\,$Q^2$ - behaviour of $F_2$ -slope ( $ \frac{\partial
F_2(x,Q^2)}{\partial \ln Q^2}$ ).
\end{enumerate}
\section{SC for $\mathbf{F_2}$}
It is well known, that $\gamma^*$ - hadron interaction goes in two
stages: (i) the transition from virtual photon to colour dipole and (ii)
the interaction of the colour dipole with the target. To illustrate how
SC work, we consider the Glauber - Mueller formula which describes the
rescatterings of the colour dipole with the target\cite{MU90}:
\begin{eqnarray} 
 F_2( x_B, Q^2 )\,\,\,&=&
\frac{N_c}{6 \pi^3} \,\sum^{N_f}_{1}\, Z^2_f\,\,\int^{Q^2}_{Q^2_0}\,\,d
Q'^2 \int d b^2_t
 \{\,\,1\,\,\,-\,\,\,e^{- \frac{1}{2}\frac{4}{9}\,\kappa (x_B,Q'^2)
\,S(b_t)}\,\,\}\,\,\nonumber\\
&+& \,\,F_2( x_B, Q^2_0 )\,,\label{GM}
\end{eqnarray}
where $S(b_t) = e^{- \frac{b^2_t}{R^2}}$ is the target  profile function
in the
impact parameter representation and $\frac{4}{9}\,\kappa (x_B,Q'^2) =
\sigma(x_B, r^2_{\perp} = 4/Q^2)$ is the cross section of the dipole
scattering in pQCD.

One can see that Eq.(~\ref{GM}) leads to
\begin{equation} \label{ASF2}
F_2(x_B,Q^2)\,\,\longrightarrow\,\,\frac{N_c}{6 \pi^3} \,\sum^{N_f}_{1}\,
Z^2_f\,\,Q^2 R^2\,\,.
\end{equation}

However, we are sure that the kinematic region of HERA is far away from
the asymptotic one. The practical calculations depend on three
ingredients: the value of $R^2$, the value of the initial virtuality
$Q^2_0$ and the initial $F_2$ at $Q^2_0$. We fix them as follows: $R^2 =
10 \,GeV^{-2}$ which corresponds to ``soft" high energy phenomenology
\cite{SOFT}, $Q^2_0 = 1 \,GeV^2$ and $F^{input}_2(x_B,Q^2_0)
=F^{GRV'94}_2(x_B,Q^2_0= 1\,GeV^2) $. Therefore, the result of calculation
should be read as ``SC for colour dipoles with the size smaller than
$r^2_{\perp}\,\leq\,4/1GeV^2 $ are equal to ..."

\begin{figure}[hbtp]
\begin{tabular} {c c }
\includegraphics[width=.5\textwidth]{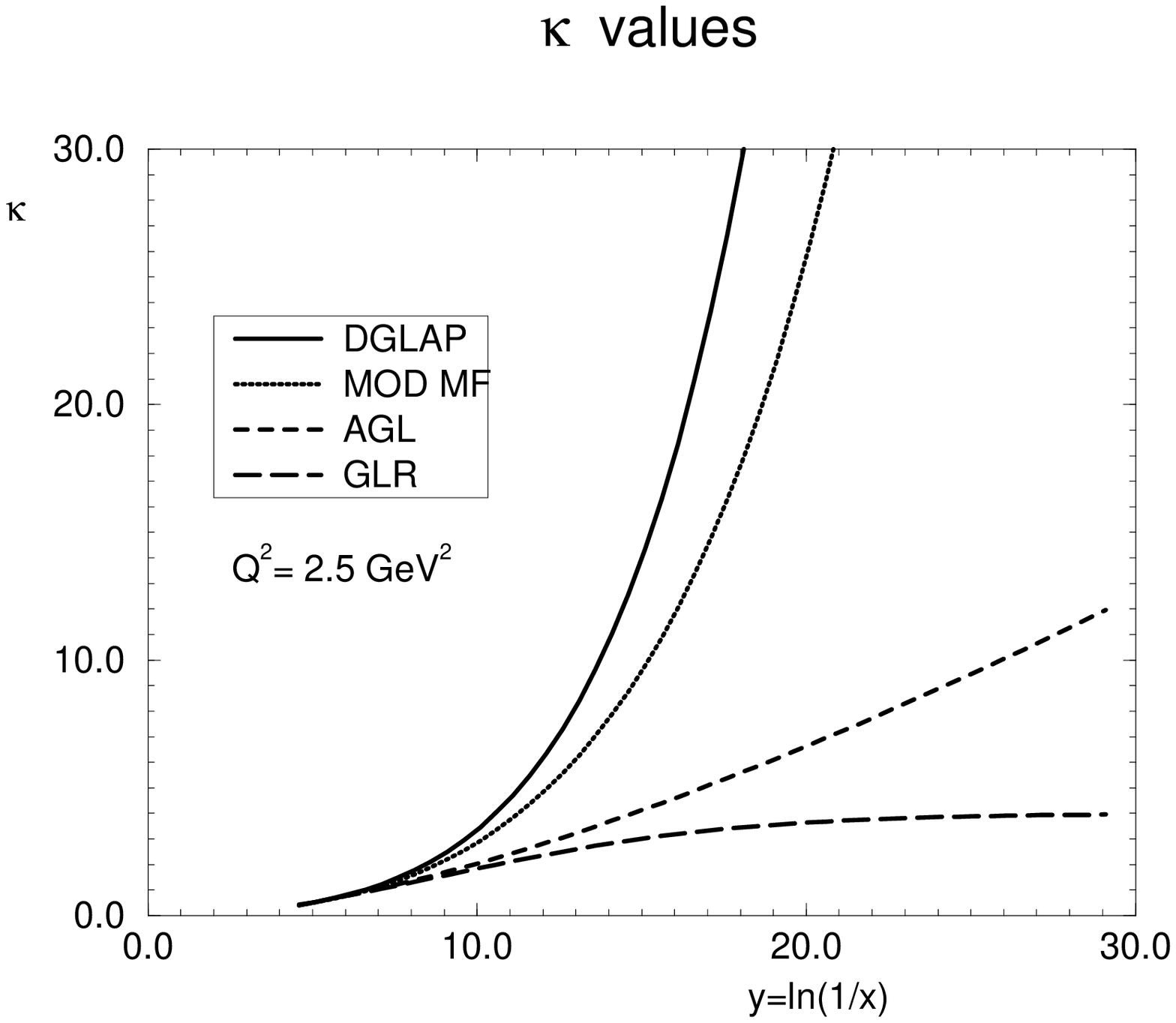} &
\includegraphics[width= 0.5\textwidth]{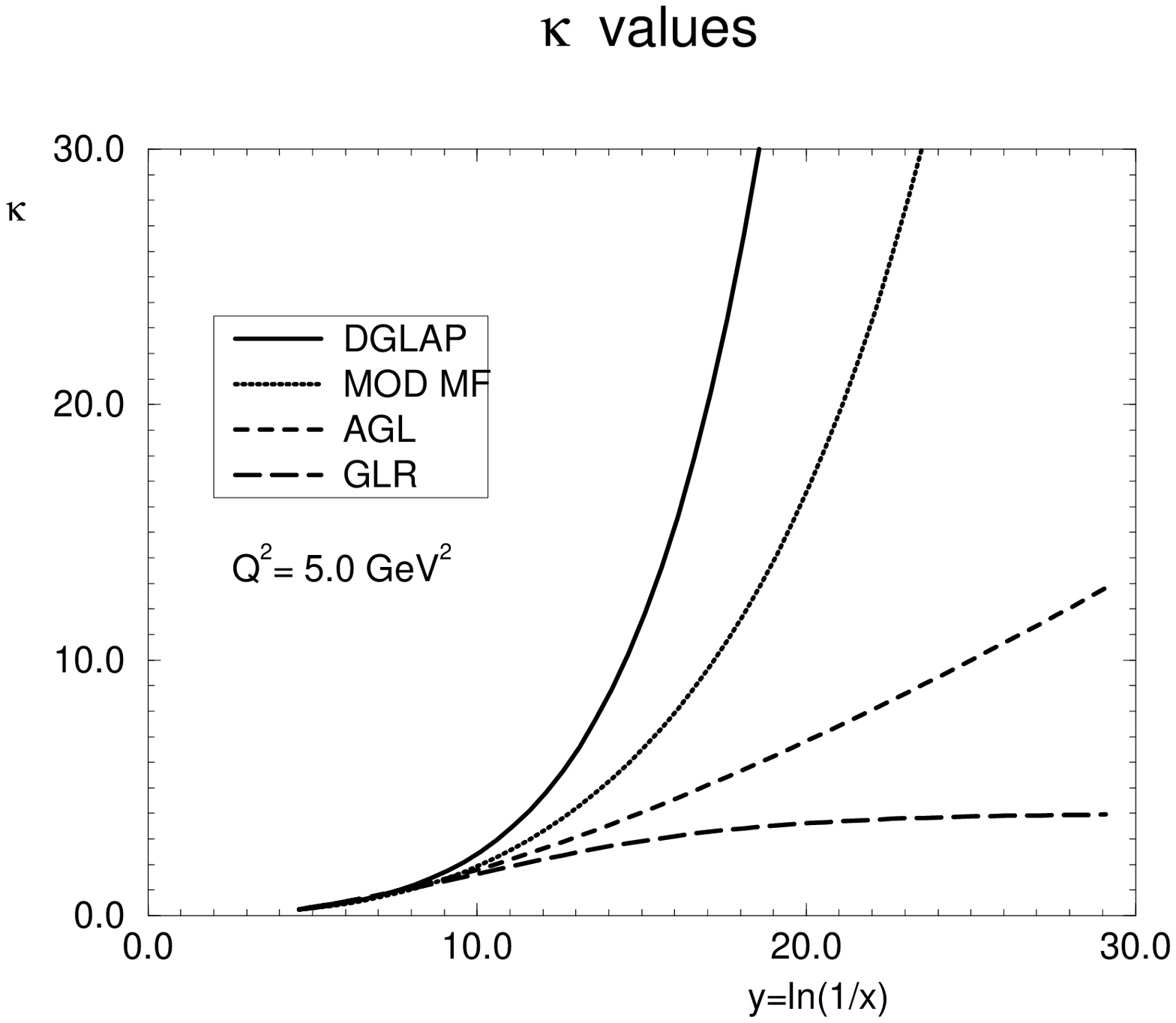}\\
\includegraphics[width=.5\textwidth]{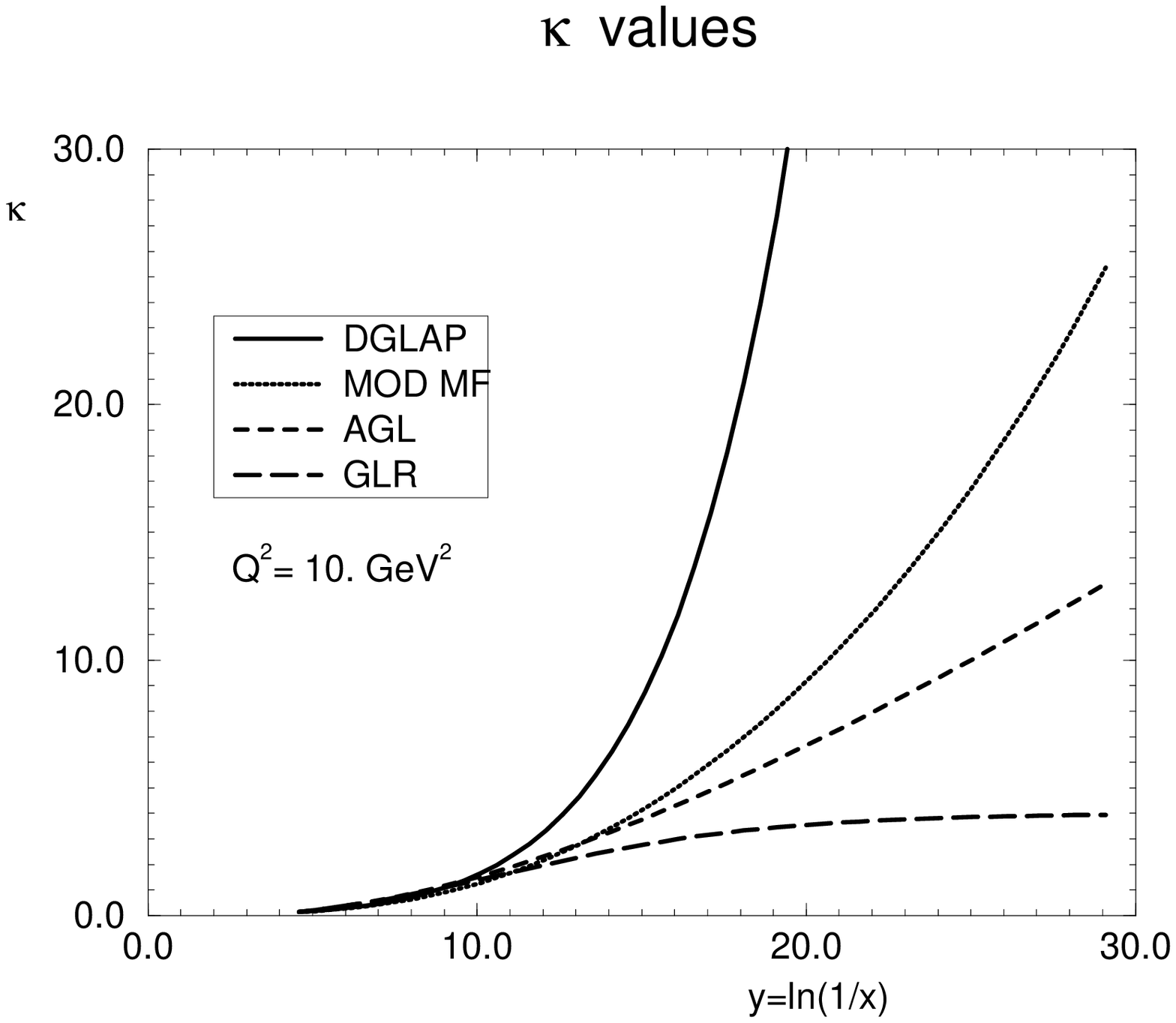} &
\includegraphics[width= 0.4\textwidth]{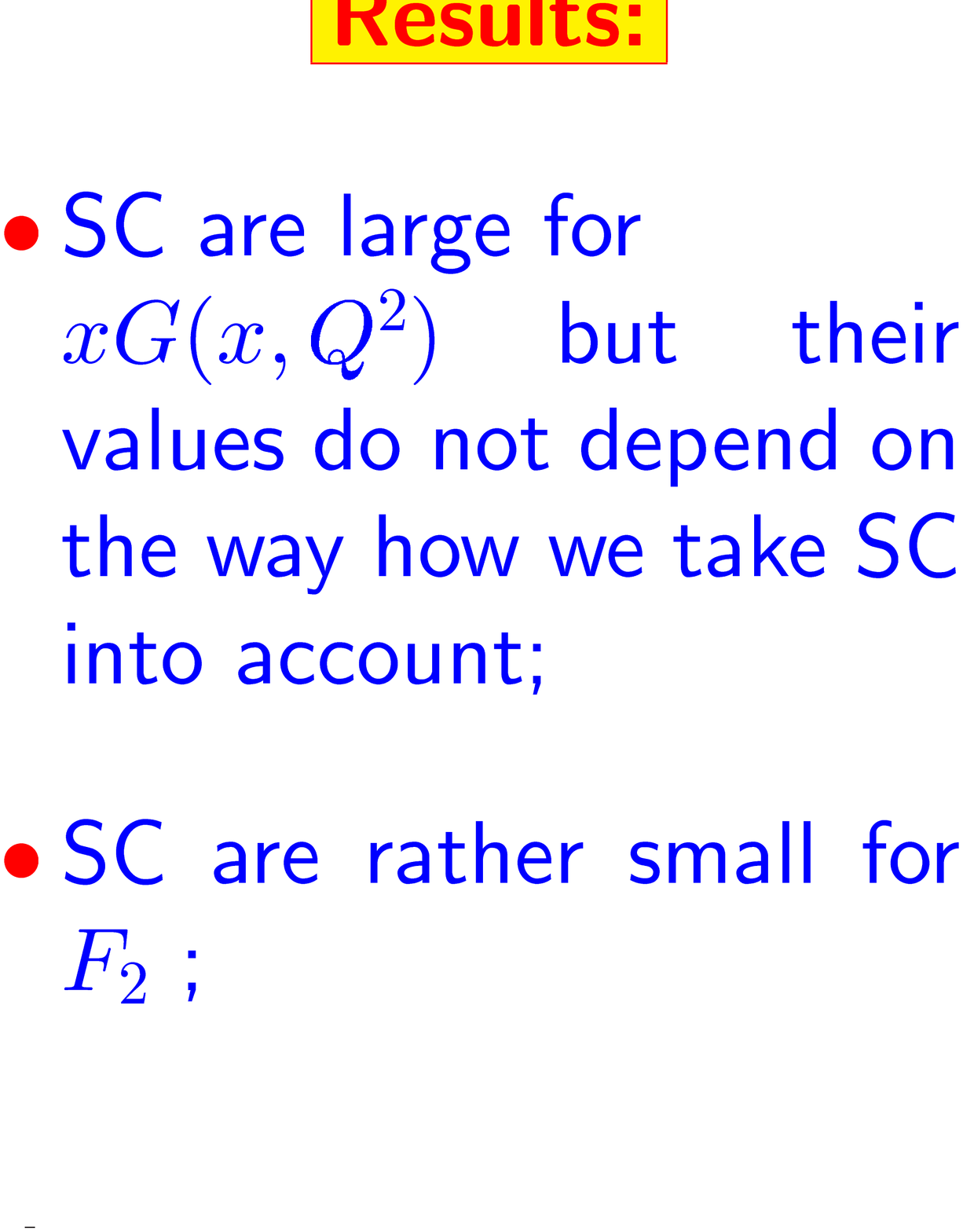}\\
\includegraphics[width=.5\textwidth]{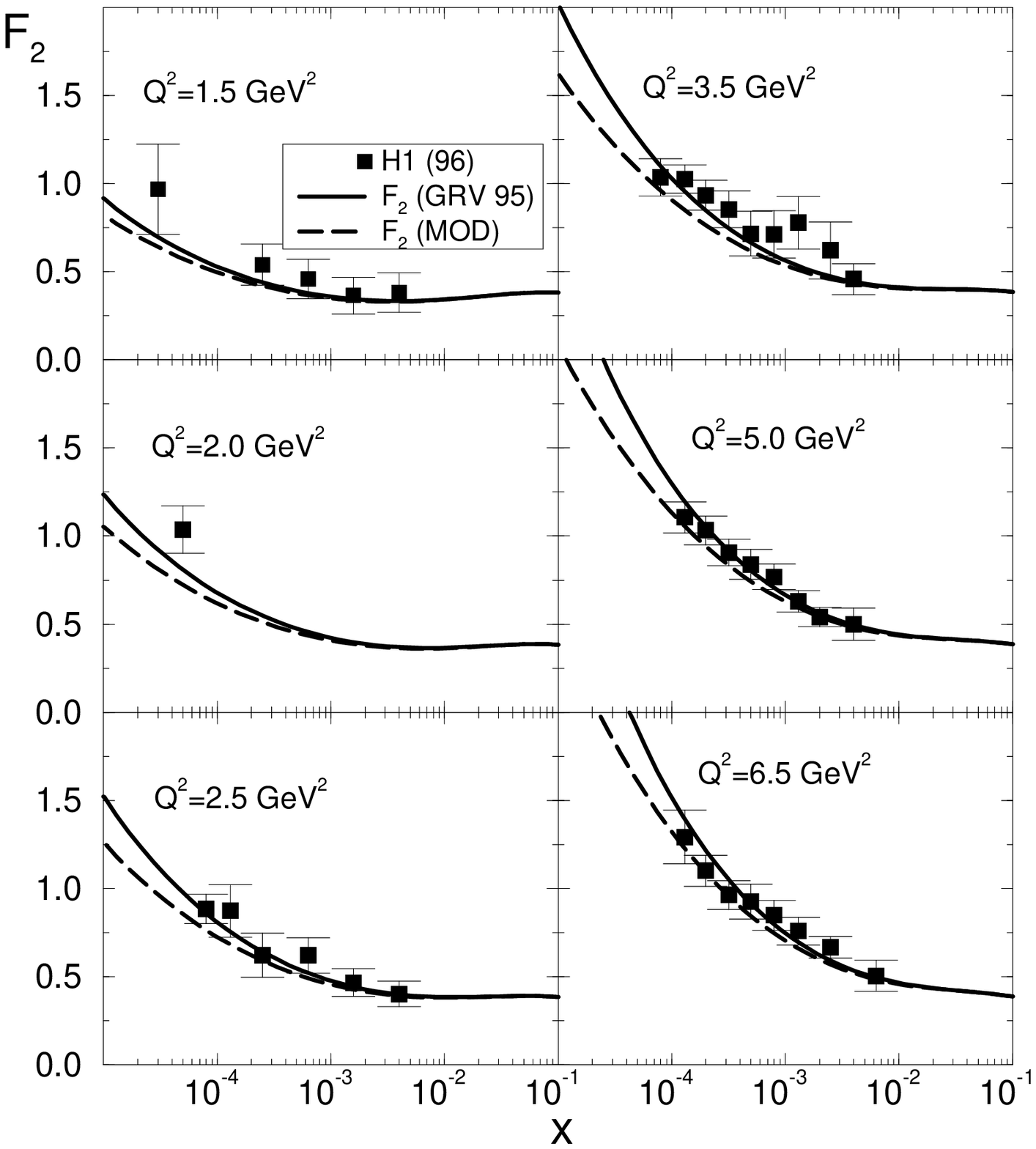} &
\includegraphics[width= 0.5\textwidth]{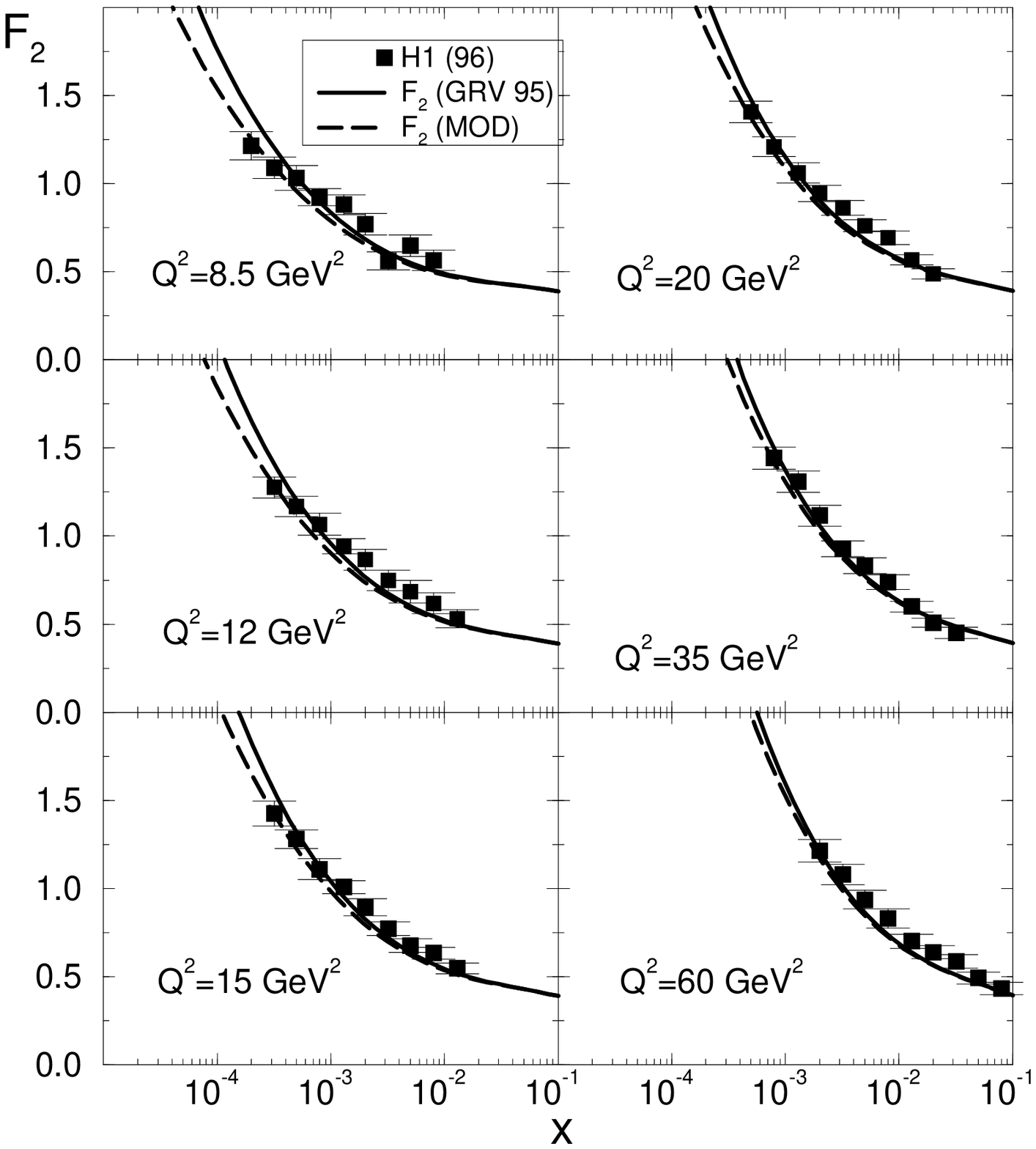}\
\end{tabular}
\caption{SC for $xG(x,Q^2)$ and $F_2$ in the HERA kinematic region .}
\end{figure}
From Fig.3 one can see that the SC are rather small for $F_2$ but they are
strong and essential for the gluon structure function. It means that we
have to look for the physics observables which will be more sensitive to
the value of the gluon structure function than $F_2$.

\section{$\mathbf{x_P}$- dependence of $\mathbf{\sigma^{DD}}$}
One of such observables is the cross section of the diffractive
dissociation and, especially, the energy dependence of this cross section. 

{\bf Data:} Both H1 and  ZEUS collaborations \cite{EXPREV} found that
\begin{equation} \label{HP}
\sigma^{DD}\,\,\propto\,\,\frac{1}{x^{2 \Delta_P}}\,\,,
\end{equation}
where $\Delta_P = \alpha_P(0) - 1 $ and the values of $\alpha_P(0)$ are:
\begin{itemize}
\item\,\,\,{\bf H1\,\,\cite{H1}\,\,\,:}\,\,\,\,\,$\alpha_P(0)$ = 1.2003
$\pm$
0.020(stat.) $\pm$ 0.013(sys)\,;
\item\,\,{\bf ZEUS\,\cite{ZEUS}\,:}\,\,$\alpha_P(0)$ = 1.1270 $\pm$
0.009(stat.) $\pm$ 0.012(sys)\,.
\end{itemize}
It is clear that the Pomeron intercept ($\alpha_P(0)$) for diffractive
processes in DIS is higher than the intercept of ``soft" Pomeron
\cite{SOFT}.

{\bf Why is it surprising and interesting?}  To answer this question we
have to recall that the cross sections for diffractive production of
quark-antiquark pair have the following form in pQCD \cite{BAWU}
\cite{GLMSM}:
\begin{eqnarray}
x_P\,\frac{d \sigma^T_{DD}( \gamma^* \,\rightarrow\, q + \bar q ) 
}{d x_P d
t}\,\,&\,\propto\,&\int^{\frac{M^2}{4}}_{Q^2_0}\,\,\frac{d
k^2_{\perp}}{k^2_{\perp}}\,\,\times\,\,\frac{\left(\,\alpha_S\, \,x_P
\,G(x_P,
\frac{k^2_{\perp}}{1 - \beta})\,\right)^2}{k^2_{\perp}} \,;\label{CR1}\\
x_P\,\frac{d \sigma^L_{DD}( \gamma^* \,\rightarrow\, q + \bar q )
}{d x_P d
t}\,\,&\,\propto\,&\,\int^{\frac{M^2}{4}}_{Q^2_0}\,\,\frac{d
k^2_{\perp}}{Q^2}\,\,\times\,\,\frac{\left(\,\alpha_S\, \,x_P
\,G(x_P,\frac{k^2_{\perp}}{1 -
\beta})\,\right)^2}{k^2_{\perp}}.\label{CR2}
\end{eqnarray}
From Eqs.(~\ref{CR1}) and (~\ref{CR2}) you can see that $k_{\perp}$
integration looks quite differently for transverse and longitudinal
polarized photon: the last one has a typical log integral over $k_{\perp}$
while the former has the integral which normally converges at small values
of $k_{\perp}$. We have the same property for production a more complex
system than $q \bar q$,  for example $ q \bar q G$ \cite{GLMSM}.
Therefore,  we expect that the diffractive production should come from
long distances where the ``soft" Pomeron contributes. However, the
experiment says a different thing, namely, that this production has a
considerable contamination from short distances. How is it possible? 
As far as we know, there is the only one explanation: SC are so strong
that $xG(x,k^2_{\perp})\,\,\propto\,\,k^2_{\perp} R^2$ (see
Eq.(~\ref{ASF2}) )  Substituting this asymptotic limit in Eq.(~\ref{CR1})
one can see that the integral becomes convergent and it sits at the upper
limit of integration which is equal to $k^2_{\perp} = Q^2_0(x)$.

Finally, we have
\begin{equation} \label{EBFIN}
x_P \,\frac{d \sigma_{DD}}{d x_P dt}
\,\,\,\longrightarrow\,\,\,\,
\left(\,x_P\,G(x_P,Q^2_0(x_P))\,\right)^2\,\,\times\,\,\frac{1}{Q^2_0(x_P)}
\end{equation}

 The calculation for $\frac{\partial xG(x,Q^2)}{\partial \ln (1/x)}$ is
given in Fig.4 for HERA kinematic region using Glauber-Mueller formula
\cite{MU90} for SC. Taking into account that $Q^2_0(x)$ in Fig.2b can be 
fitted as 
$ Q^2_0(x)\,\,\,=\,\,1\,\,GeV^2\,\left(\,\frac{x}{x_0}\,\right)^{-
\lambda}$ with $\lambda = 0.54$ and $x_0 = 10^{-2}$ we see from
Eq.(~\ref{EBFIN}) and Fig.4 that we are able to reproduce the experimental
value of $\alpha_P(0)$ and conclude that the typical $k^2_{\perp}$ which
are dominant in the integral is not small ( $ k^2_{\perp}\,\,\approx\,\,1 -
2 \,GeV^2$ \cite{GLMSM} ). For Golec-Biernat and W\"{u}sthoff approach,
which
we will discuss late, $\lambda = 0.288$ and the value of typical
$k^2_{\perp}$ turns out to be higher.

\begin{figure}[hbtp]
\begin{center}
\includegraphics[width=0.8\textwidth]{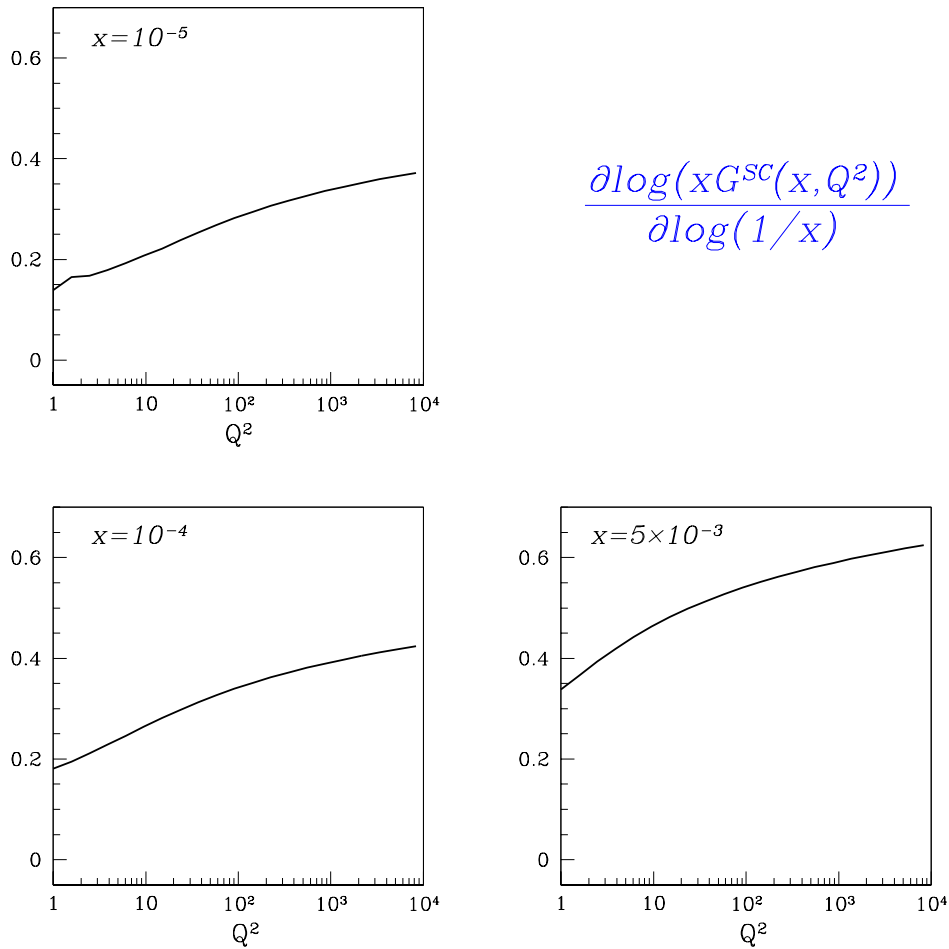}
\end{center}
\caption[]{Effective Pomeron intercept for gluon structure
function calculated using Glauber-Mueller formula for SC}
\label{eps1.5}  
\end{figure}

\section{The $\mathbf{ Q^2}$-
dependence of the $\mathbf{F_2}$ - slope. }

{~~~~}{\bf Data:} The experimental data \cite{ZEUS} for $F_2$ - slope $
d
F_2(x,Q^2)/d \ln Q^2$ are shown in Fig.5a (Caldwell plot). These data give
rise to a hope that the matching between ``hard" ( short distance)  and
``soft" (long distance) processes occurs at sufficiently large $Q^2$ since
 the $F_2$-slope starts to deviate from the DGLAP predictions around $Q^2
\,\approx\,5 - 8\,GeV^2$.

{\bf $\mathbf{F_2}$-slope and SC:}  Our principle idea, as we have
mentioned in the beginning of the talk,  is that matching between ``hard"
and ``soft" processes is provided by the  hdQCD phase in the parton
cascade or, in other words, due to strong SC. The asymptotic behaviour of
$F_2\,\propto\,Q^2 R^2 $ for $ Q^2 \,\leq\,Q^2_0(x)$ leads to $d
F_2(x,Q^2)/d \ln Q^2 \,\propto\,Q^2 R^2$ at $ Q^2 \,\leq\,Q^2_0(x)$
 (see Eq.(~\ref{ASF2}) ) and this behaviour  supports our
point of view \cite{GLMSLOPE}\cite{MUSLOPE}. 

However, we have two problems
to solve before making any conclusion: (i) the experimental data are taken
at different points  $(x,Q^2)$ and therefore could be interpreted as the
change of $x$-behaviour rather than $Q^2$ one; and (ii)   the value
of $F_2$-slope is quite different from the value of $F_2$ while for the
asymptotic solution it should be the same.  Therefore, we have to
calculate $F_2$ - slope to understand them. The result of calculation
using the Glauber-Mueller formula \cite{GLMSLOPE} is presented in Fig.5b.
One can see that (i) the experimental data shows rather $Q^2$ - behaviour
then the $x$-dependence, which is not  qualitatively influenced by
SC;
and (ii) SC are able to describe both the value and the $Q^2$-behaviour of
the experimental data. Fig.5b shows also that the ALLM'97 parameterization
\cite{ALLM}, which can be viewed as the phenomenological description of
the
experimental data,  has the same features as our calculation confirming
the
fact that the data show $Q^2$ - dependence but not $x$-behaviour of the
$F_2$-slope.

\begin{figure}[hbtp]
\begin{tabular} {c c }
\includegraphics[width=.4\textwidth]{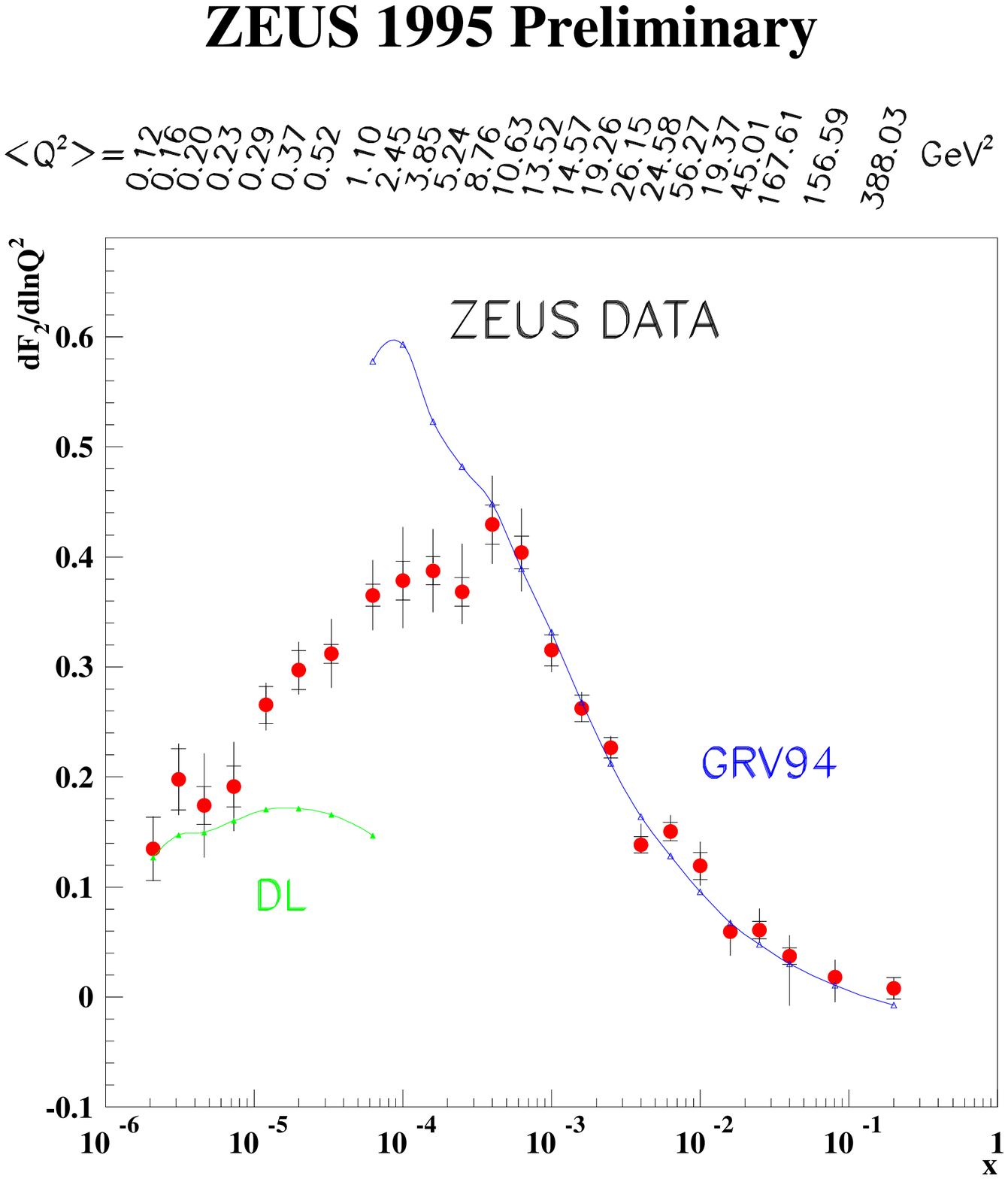} &
\includegraphics[width= 0.5\textwidth]{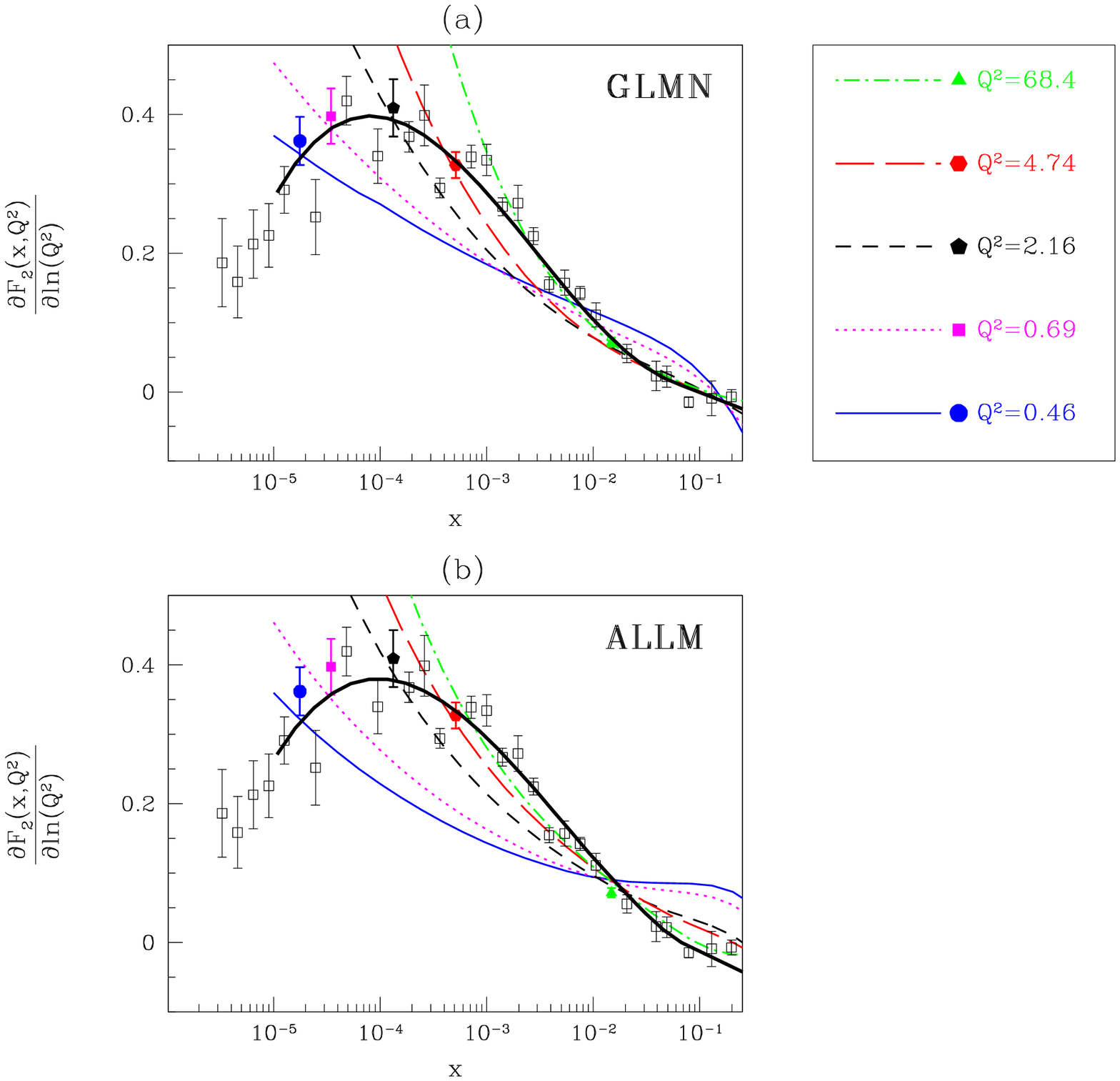}\\
Fig.5a & Fig.5b
\end{tabular}
\caption{$F_2$-slope: experimental data (Caldwell plot ) 
(Fig.5a) and calculations using Glauber-Mueller formula (Fig.5b)
.}
\end{figure}

\section{ Golec-Biernat and W\"{u}sthoff  Approach}
Golec-Biernat and W\"{u}sthoff \cite{GW} suggested a phenomenological
approach which takes into account the key idea of hdQCD, namely, the new
scale of hardness in the parton cascade. They use for $\gamma^* p$ cross
section the following formula \cite{GW}
\begin{eqnarray}
\sigma_{tot} ( \gamma^* p )\,& = &\, \int d^2 r_{\perp}
\int^1_0 \,d z \,| \Psi( Q^2; r_{\perp},z ) |^2 \,\,\sigma_{tot} (
r^2_{\perp}, x)\,\,;\label{GW1}\\
\sigma(x,r_{\perp})\,&=&\,\sigma_0\,\,\{\,\,1\,\,\,-\,\,\,e^{-
\,\frac{
r^2_{\perp}}{R^2(x)}}\,\,\}\,\,;\label{GW2}\\
R^2(x)\,&=&\,1/Q^2_0(x)\,\, \,\,with\,\,\,
Q^2_0(x)\,\,=\,\,Q^2_0 \,\left(\,\frac{x}{x_0}\,\right)^{- \lambda}\,\,.
\label{GW3}
\end{eqnarray}

Extracting  parameters of their model from fitting of the experimental
data, namely, $\sigma_0 $=23.03 mb,
$\lambda$ =  0.288, $Q^2_0 = 1 GeV^2$  and 
$x_0$ = 3.04\,$10^{-4}$ , they described quite well all data on total and
diffractive cross sections in DIS (see Fig.6 ).

\begin{figure}[hbtp]
\begin{tabular} {c c }
\includegraphics[width=.55\textwidth]{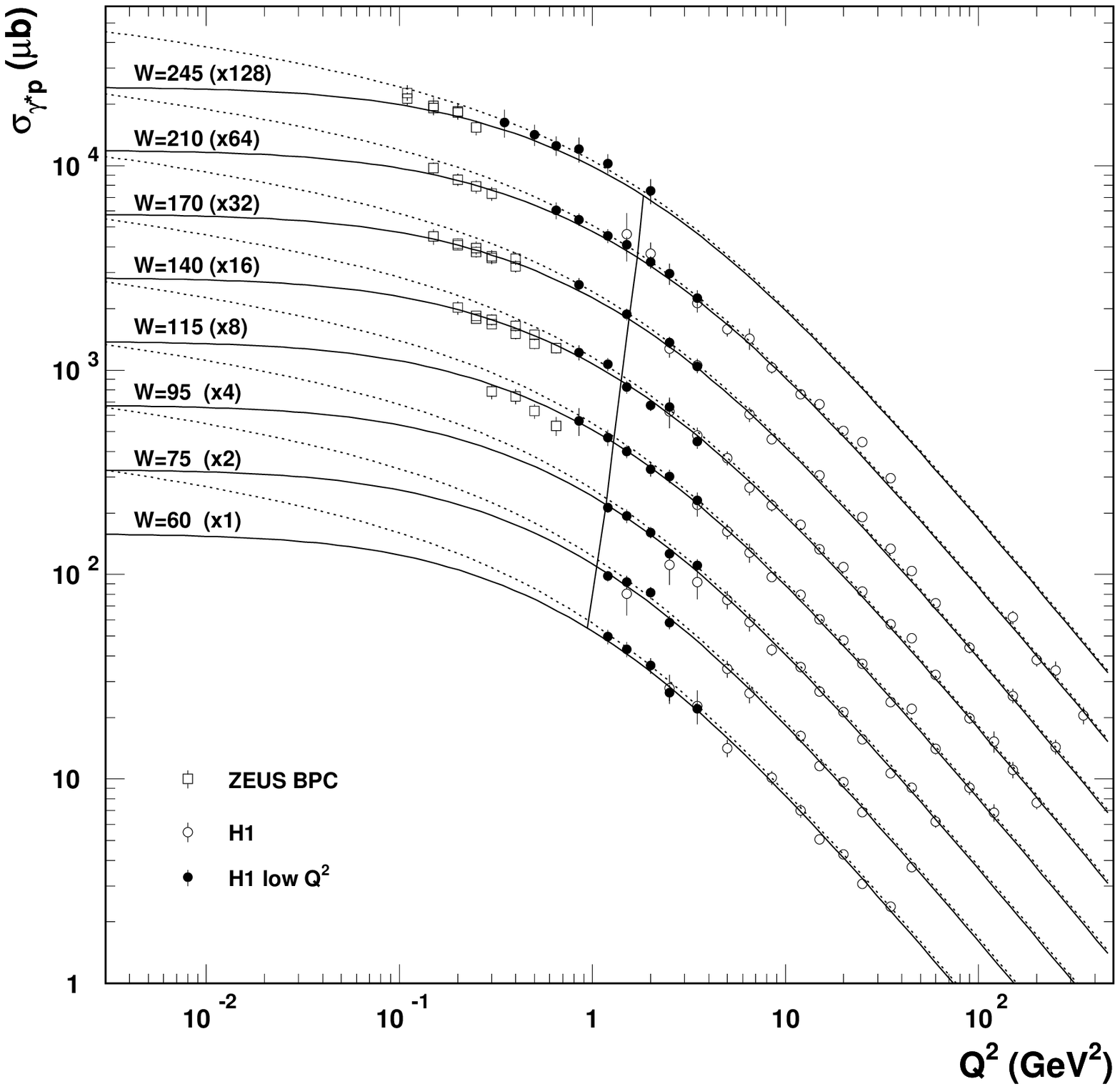} &
\includegraphics[width= 0.45\textwidth]{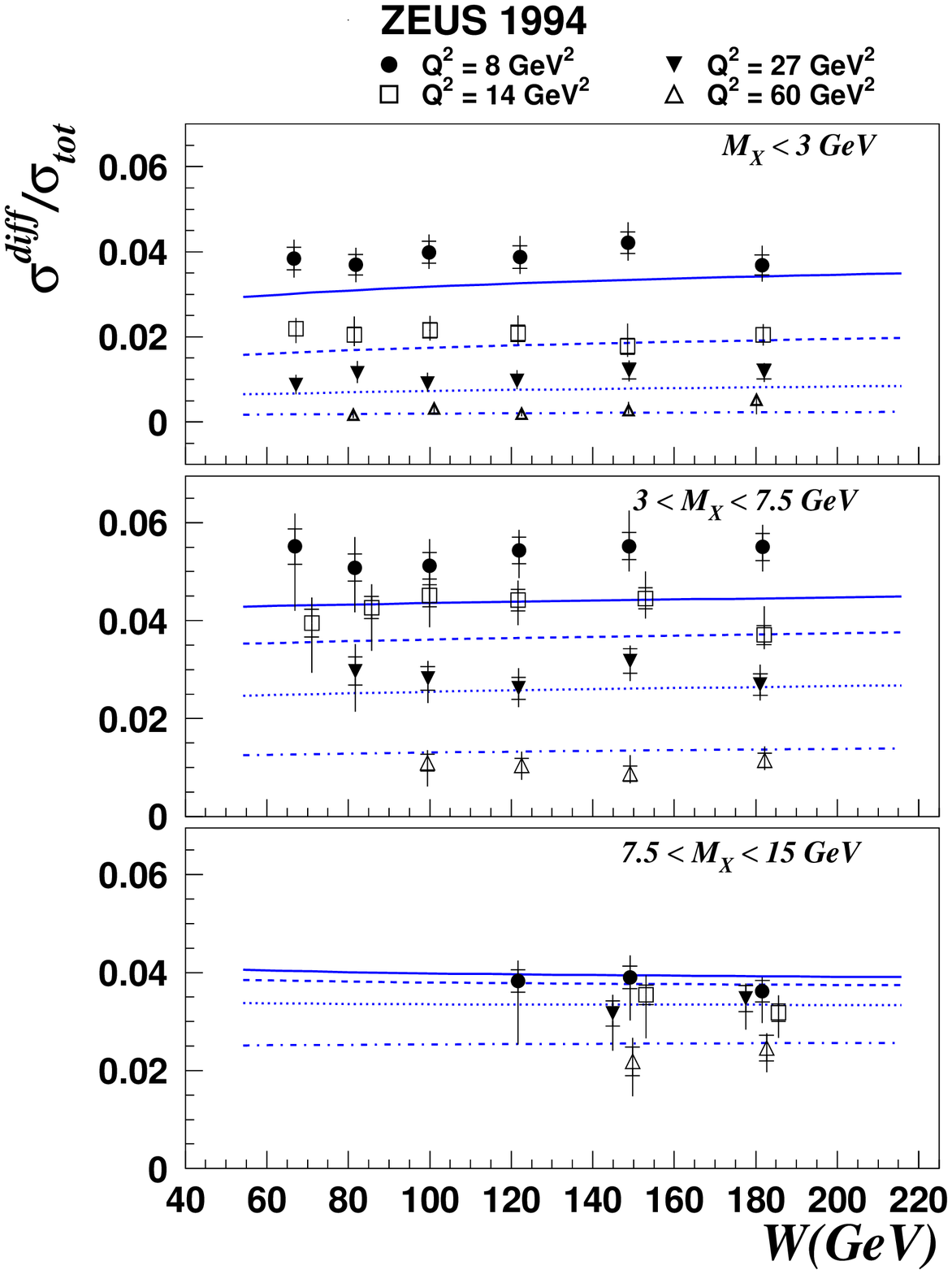}\\
Fig.6a & Fig.6b
\end{tabular}
\caption{$\sigma_{tot}(\gamma^*p)$ amnd the ration
$\sigma^{DD}/\sigma_{tot}$ for DIS in Golec-Biernat and W\"{u}sthoff
model. Vertical line in Fig.6a is the $Q^2_0(x)$ given by Eq.(16).}
 \end{figure}
\section{Why Have We Only Indications?}
The  answer is: because we have or can have an alternative
explanation of each separate fact. For example, we can describe the $F_2$
-slope behaviour changing the initial $x$-distribution for the DGLAP
evolution equations \cite{MRST}. Our difficulties in an interpretation of
the experimental data is seen  in Fig.6a where the new scale $Q^2_0(x)$ is
plotted. One can see that $Q^2_0(x)$  is almost constant in HERA kinematic
region. It means that we can put the initial condition for the evolution
equation at $Q^2_0 = < Q^2_0(x)>$ where $< Q^2_0(x)>$ is the average new
scale in HERA region. Therefore, SC can be absorbed to large extent in the
initial condition and the question, that can and should be asked, is how   
well motivated these conditions are. For example, I do not think that
initial gluon distribution in MRST parameterization \cite{MRST}, needed to
describe the $F_2$ - slope data, can be considered as a natural one.
\section{Summary}
We hope we convinced you that (i) hdQCD is in a good theoretical shape;
(ii) the hdQCD region has been reached at HERA; (iii) HERA data do not
contradict the strong SC effects; (iv) there are at least two indications
on SC effects in HERA data: $Q^2$ behaviour of $F_2$ slope and $x_P$
behaviour of diffractive cross section in DIS; and (v) the HERA data and
the hdQCD theory gave an impetus for a very successful phenomenology for
matching ``hard" and ``soft" physics.

We would like to finish this talk with rather long citation:
``Small $x$ Physics is still in its infancy. Its relations
to heavy ion physics, mathematical physics and soft hadron physics along
with a rich variety of possible signature makes it central
for QCD studies over the next decade"
( A.H. Mueller, B. M\"{u}ller, G. Rebbi and W.H. Smith
``Report of the DPF Long Range Planning WG on QCD"). Hopefully, we will
learn more on low $x$ physics by the next Ringberg Workshop.

\end{document}